\documentclass[a4paper]{jpconf}
\usepackage{graphicx}
\begin{document}
\newcommand{\dis}[1]{\begin{equation}\begin{split}#1\end{split}\end{equation}}

\newcommand{\tev}{\,\textrm{TeV}}
\newcommand{\gev}{\,\textrm{GeV}}
\newcommand{\meV}{\,\mathrm{MeV}}
\newcommand{\keV}{\,\mathrm{keV}}
\newcommand{\eV}{\,\mathrm{eV}}
\newcommand{\NDMMSSM}{$N_{\textrm{DM}}$MSSM}
\newcommand{\NPQ}{$N_{\textrm{PQ}}$MSSM}
\newcommand{\MG}{{$M_{\rm GUT}$}}
\newcommand{\ie}{{\it i.e.}\ }

\def\E6{{\rm E_6}}
\def\EE8{{\rm E_8\times E_8}}
\def\fourb{\overline{\bf 4}\,}
\def\four{{\bf 4}}
\def\two{{\bf 2}}
\def\fiveb{\overline{\bf 5}\,}
\def\five{{\bf 5}}
\def\tenb{\overline{\bf 10}\,}
\def\ten{{\bf 10}}
\def\one{{\bf 1}}
\def\six{{\bf 6}}
\def\threeb{\overline{\bf 3}\,}
\def\three{{\bf 3}}
\def\fif{{\bf 15}}
\def\sixb{\overline{\bf 6}\,}
\def\six{{\bf 6}}

\title{Effective SUSY modified after LHC run\footnote{Talk presented at GUT 2012, Kyoto, Japan 15-17 March 2012, and PASCOS 2012, Merida, Mexico, 3-8 June 2012.}}

\author{Jihn E. Kim}

\address{Department of Physics, Seoul National University, Seoul 151-747, Korea, and\\
 GIST College, Gwangju Institute of Science and Technology, Gwangju 500-712, Korea
}
\ead{jihnekim@gist.ac.kr}

\begin{abstract}
Some supersymmetric models after the recent LHC reports are discussed. Especially, the light Higgs boson around 125\,GeV is attempted to be accommodated. An extra $Z'$ from GUTs, and the scalar and pseudoscalar Higgs bosons in supersymmetric extension of the standard model are also discussed.
\end{abstract}


\section{Introduction}
I will discuss a beyond standard model(BSM) view based on my recent works.

The most recent interest in high energy physics is the LHC reports on the Higgs boson discovery at the mass range of 125  GeV \cite{CERNJuly4}. This small Higgs boson mass compared to the Planck mass needs a huge hierarchy of mass scales, inviting solutions of the hierarchy problem. The attempt to explain the Higgs boson mass within one GeV limit has been possible in supersymmetry(SUSY) extension of the standard model(SM). But the LHC data is not consistent with the constrained minimal supersymmetric standard model(CMSSM) prediction in the region $ m_{\rm gluino}m_{\rm squark}< 1\tev^2$. A small Higgs boson mass ($m_h\simeq 0.125\,\tev$) needs a large stop mass or/and a large $A$-term in the CMSSM.

Of course, there have been several important issues to be considered in the BSM: a new CP, axions, SUSY, string, etc. For example, if there exists a new force around the electroweak scale, an extra $Z'$ is expected at the TeV scale. As done for the $Z$ boson \cite{Glashow61}, the indirect limits for the extra force have been studied for a long time. For example, some string inspired $Z'$ mass limit has been given as 1.36 TeV \cite{Earler09}. Recently, the LHC data seems to give limits on the $Z'$ mass limit in the SUSY SM(SSM) as 2.2 TeV \cite{AtlasKyoto}.

For the extra  $Z'$, the first interest resides in whether it can arise from a GUT or not. For GUT groups, we consider a few interesting simple and semi-simple groups such as SU(5), SO(10), SU(6), trinification SU(3)$^3$. These are subgroups of $\E6$. Therefore, the $Z'$ issue of those interesting GUT groups is contained in the study of $\E6$. Then, it is sufficient to study the diagonal generators of $\E6$, because any Cartan subalgebra generator of SO(10), SU(6), and trinification SU(3)$^3$ can be expressed as a linear combination of the $\E6$ diagonal generators. With this philosophy, we consider any rank 6 subgroup of $\E6$, and SU(6)$\times$SU(2) is particularly useful for this purpose because one can see all the quantum numbers of representations as displayed below:
\begin{eqnarray}
 (\fif,\one) &\equiv& \left(\begin{array}{cccccc}
0&u^c& -u^c &u&d&D \\ -u^c&0&u^c & u&d&D \\
u^c&-u^c&0 &u&d&D \\ -u&-u&-u&0 & e^c&H_u^{+} \\
-d&-d&-d& -e^c&0 & H_u^{0} \\
-D&-D&-D&-H_u^{+}  & -H_u^{0} &0 \\
\end{array}\right),\nonumber\\[0.5em]
 (\sixb,\two^\uparrow) &\equiv& \left(\begin{array}{c}
  {d^c}\\ { d^c}\\ { d^c} \\  -\nu_{e}  \\ {e} \\  { N}
\end{array}\right),~~~~
  (\sixb,\two^\downarrow)=\left(\begin{array}{c}
  {D^c}\\  {D^c}\\  {D^c}\\   -H_d^0   \\  H_d^- \\  N'
\end{array}\right).\label{eq:SUsixGUT}
\end{eqnarray}
The vertical (SU(6)) and the horizontal (SU(2)) directions of the $\fif$ and $\sixb$ introduce six diagonal generators of $\E6$.

Firstly, we note that baryon number $B$ cannot be a U(1) generator of $\E6$ as discussed by many people and as shown recently in \cite{KimShin12}.
If it is not baryon number, this U(1)$'$ may be constructed such that  $Z'$ does not couple to leptons, which has been known as a leptophobic U(1)$'$.  Considering  $Z'$ from $\E6$ subgroups, the $\rho$ parameter of the neutral current study from the LEPII data does not allow the  $Z'$ mass below 10 TeV \cite{KimShin12}. But if we do not restrict to a subgroup of $\E6$, a TeV scale $Z'$ is perfectly allowed from string, which is because the SSM in this case is basically obtained from $\EE8'$, \ie beyond Eq. (\ref{eq:SUsixGUT}) of $\E6$.
Still the extra $Z'$ with a large $Z'$ mass from $\E6$ can contribute to low energy physics. In SUSY models, they can work as the messengers of SUSY breaking sector to the SSM physics at TeV scale.

\section{Effective SUSY}
Effective SUSY has been proposed in the middle of 1990s. In this regard, we note that the third generation and Higgs doublets are the necessary ingredients of the gauge hierarchy solution through SUSY \cite{effSUSY95}. It is also useful to fit the $(g-2)_{\mu}$ data \cite{gminus2effSUSY}.

To remove too large FCNC effects among the light two family members by raising the 1st and 2nd family squark masses a little bit, one needs a symmetry among the first two family squarks such as a U(2) symmetry. But, if the 1st and 2nd family squark masses are raised above $10^{5-6}$ GeV, one does not need such a symmetry among the light family members.

We heard here on the LHC results interpreted in the constrained MSSM and hence a simplified SUSY model \cite{AtlasKyoto}. The latest CMS results from Razor analysis at the 4.4 fb$^{-1}$ level excludes squark and gluino masses up to 1.35\,TeV, and the
latest ATLAS results from $m_{\rm eff}$ on 0-lepton at 4.7\,fb$^{-1}$  is mapped on with the massless LSP assumption on $m_{\rm gluino}<940\,{\rm GeV}$ and $ m_{\rm squark}<1380\,{\rm GeV}$. For 0-lepton + high multiplicity jets ($\ge 6$ to $\ge 9$) gave $m_{\rm gluino}<880\,{\rm GeV}$. So, squark masses seems to be heavier than TeV in a simplified MSSM. But, if this simplified assumption is not taken, these limits do not apply. One attractive such case is effective SUSY(effSUSY).
\vskip 0.2cm
\noindent {\bf $Z{\,}'$ mediation:}
The effSUSY was a phenomenological condition for the squark and gluino mass spectra \cite{effSUSY95}. For a long time a theoretical model has not been proposed.  One needs a model distinguishing families. For example, in the gravity mediation scenario, one needs to assign specifically the first two family and the third family members in topologically different manifolds. This needs a complete model in the ultra-violet completion, e.g. in string compactification in the bulk and different fixed points. So, in the gravity mediation, one needs an explicit string model.

In the gauge mediated SUSY breaking(GMSB) scenario, one can pursue the effSUSY at the field theory level since the gravity mediation is assumed to be sub-dominant. However, at present there has not appeared a serious GMSB model. If an SSM is obtained from string compactification, the group nature is well worked out from the $\EE8'$ heterotic string. Suppose that the hidden sector is SU(4)$'$ or SU(5)$'$ which are favored most. For SU(4)$'$, an anomaly-free chiral representation needs a gigantic structure. For SU(5)$'$, however, it is easy to obtain an anomaly-free chiral representation, for example $\ten' +\fiveb'$. If we include $\ten' +\fiveb'$, there should be 45 visible sector fermions and at least 15 (from $\ten' +\fiveb'$) hidden sector fermions. These require at least  60 chiral fermions from string compactification. In addition, we need the messenger sector fermions carrying the SU(5)$'$ charge. Namely, the messenger sector in a general GMSB must carry color, weak and U(1)$_Y$ charges in addition to the hidden sector gauge charge.  For $\five' +\fiveb'$ messenger, the minimum number of the messengers are therefore 50 (ten times color 3 and weak 2). Thus, at least we need 110 chiral fields (with the DSB sector $\ten' +\fiveb'$ and the messenger sector $\five' +\fiveb'$) from the string  compactification. Experence tells us that roughly 150 fermions appear in the spectrum from string compactification. So, we notice that the GMSB scenario needs more chiral fields to realize a complete mediation scheme, which tightens the model building possibilities from string.

One interesting GMSB scenario is the $Z{\,}'$-mediation where SUSY breaking at the hidden sector is manifested in the visible sector (SM) by the U(1)$'$ charged fields. If the messenger sector has only $\five' +\fiveb'$, we need only 10 instead of 50 of the preceding paragraph. The $Z{\,}'$-mediation has been known for some time \cite{MohapatraZp}, but the recent suggestion by Langacker {\it et al.} is relevant for this talk \cite{LangZp08}.
Furthermore, we find that it can be obtained from a string compactification \cite{Kimstable07}.

Langacker {\it et al.} discuss mediation with one  $Z'$ and Mohapatra and Nandi  \cite{MohapatraZp} discuss mediation through  $Z'$ and $Z$. In the latter case,  it is proper to  call it the mixed mediation (MM). In our study, we find that the MM needs a fine-tuning of parameters \cite{JeongKimSeo}. So, we discuss $Z'$ mediation the type considered by Langacker {\it et al.} \cite{LangZp08}. However, ours is different from that of  Langacker {\it et al.} in that we consider a family-dependent $Z'$ mediation while  Langacker {\it et al.} considered the family-universal $Z'$ mediation  \cite{LangZp08}.
\vskip 0.2cm
\noindent {\bf Effective SUSY:}

The U(1)$'$ charge is carried by the mediators and the first two family members, but
NOT by the 3rd family members and the Higgs doublets. Because the U(1)$'$ charge is the differentiating one between the first two and the third family members, it is a CALCULABLE MODEL with only a SUSY gauge theory knowledge. In particular,  we need a light Higgs doublets pair for the hierarchy solution via SUSY. The SU(2)$\times$U(1) breaking is achieved radiatively as pointed out in \cite{Ibanez82}. The spectrum of $H_u, H_d$ and the 3rd family members being light is called effSUSY or some other words \cite{effSUSY95}. [Note added after the talk: After the discovery of the Higgs boson \cite{CERNJuly4}, the opposite view `Inverted effective SUSY' where the 3rd family members are considered to be the heaviest family members has been proposed \cite{IeffSUSY12}.]

In this regard, we note that the experimental limits on Higgs boson searches from the ATLAS and CMS data give the light Higgs boson mass around 125 GeV if it is light \cite{Earler12}. Here, the combined constraint from ATLAS and CMS is given by Jinnouchi \cite{AtlasKyoto}.

\begin{figure}[h!]
  \begin{center}
  \begin{tabular}{c}
   \includegraphics[width=0.35\textwidth]{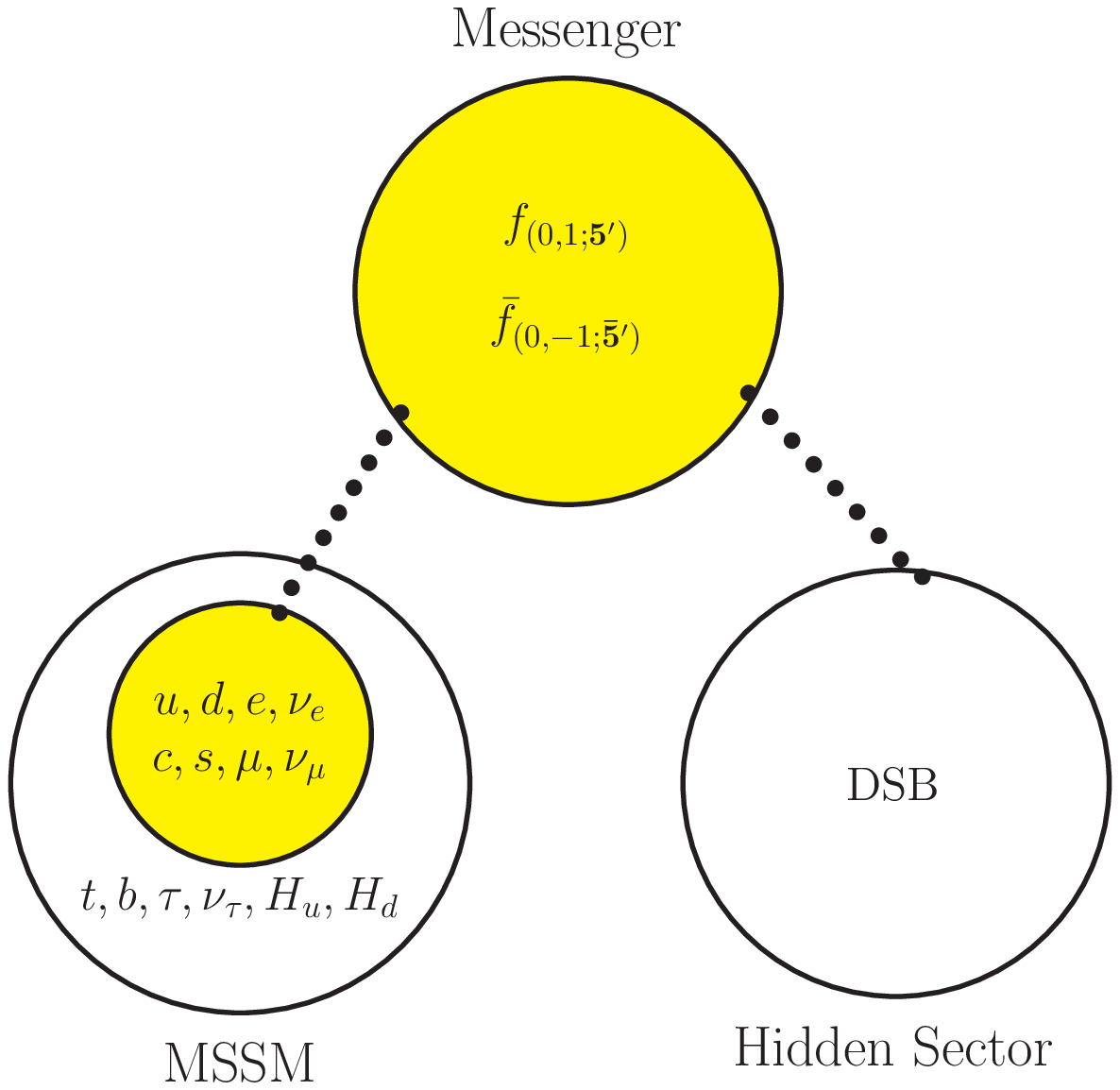}\hskip 1cm \includegraphics[width=0.45\textwidth]{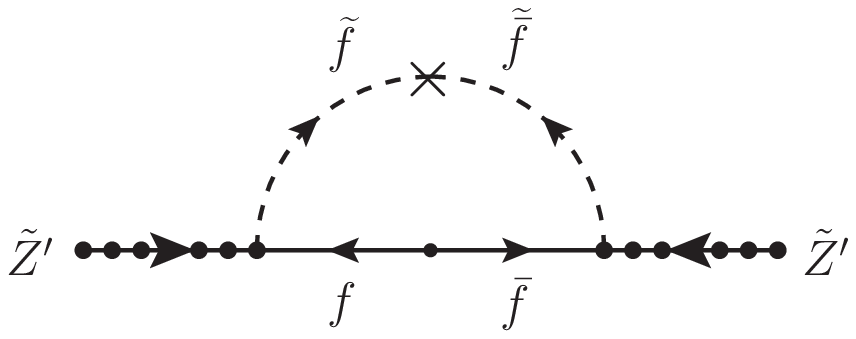}
   \end{tabular}
  \end{center}
  \caption{An effSUSY through the $Z{\,}'$ mediation  and the mass diagram of Zprimino.  The $Z{\,}'$ is the bulletted dots and the Zprimino is the bulleted line. The SUSY breaking insertion from DSB is $\times$. This soft mass shown as the Feynman diagram is added to the SUSY mass.}
\label{fig:effSUSYZprime}
\end{figure}

The idea of $Z{\,}'$ mediation toward effSUSY is pictorially depicted in Fig. \ref{fig:effSUSYZprime}.
Here, the superpartner of $Z{\,}'$, \ie Zprimino  $\tilde Z{\,}'$ is the one mediating SUSY breaking. The SUSY breaking source is shown as dynamical SUSY braking(DSB), probably a hidden-color sector. The DSB matter does not carry the weak hypercharge $Y$. The messenger sector carries the DSB hidden-color and also the $Z{\,}'$ charge but does not carry the weak hypercharge $Y$. The first two families carry  the $Z{\,}'$ charge but the third family does not carry  the $Z{\,}'$ charge. Finally, the Higgs fields do not carry the $Z{\,}'$ charge. To obtain this kind, we assign different quantum numbers to the third family members from those of the first two families.

The mediation mechanism is called the $Z{\,}'$ meditation but ours is different from that of Ref. \cite{LangZp08} in that here we need family dependent $Z{\,}'$ charges while that of Ref. \cite{LangZp08} introduces the family universal $Z{\,}'$ charges.

The largest SUSY splitting mass is the one arising in the Zprimino mass as shown as the Feynman diagram in Fig. \ref{fig:effSUSYZprime}. The SUSY splitting mass is
\begin{eqnarray}
\Delta M_{\tilde Z'}\simeq \frac{g_{Y'^{\,2}}}{16\pi^2}\frac{F_{\rm mess}'}{M_{\rm mess}'}
\end{eqnarray}

The next largest splitting appears for the first two family sfermions which appear  at one-loop level beyond $\Delta M_{Z'}$,
\begin{figure}[h!]
  \begin{center}
  \begin{tabular}{c}
   \includegraphics[width=0.45\textwidth]{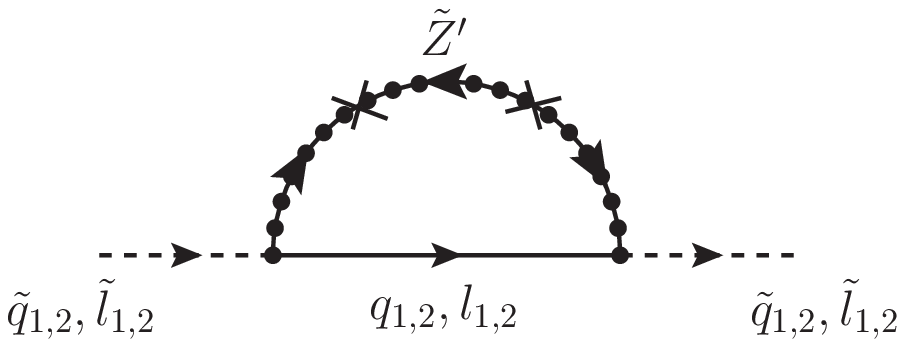}\hskip 1cm
    \includegraphics[width=0.37\textwidth]{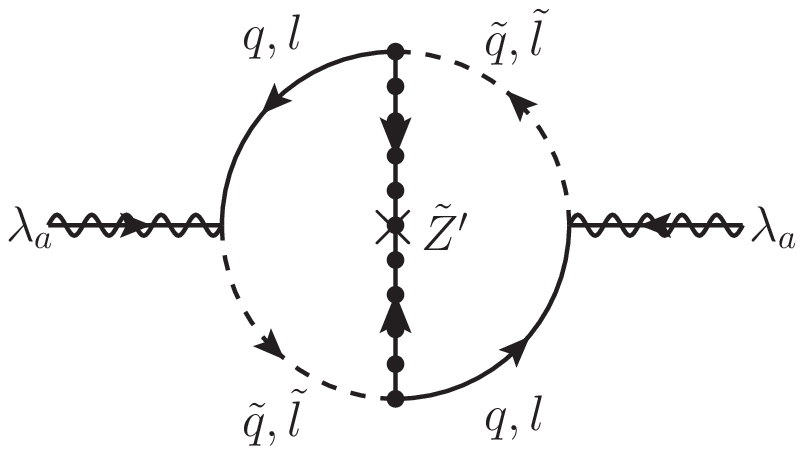}
   \end{tabular}
  \end{center}
  \caption{The first two family sfermion($\tilde{q}_{1,2},\tilde{l}_{1,2}$) mass diagrams(left panel) and the mass diagram of the SM gauginos(right panel). The SUSY breaking from Zprimino sector is shown as $\times$.  The $\tilde{Z}'$ line is a bulleted line.}
\label{fig:OneTwoFamgaugino}
\end{figure}
as shown in the left panel of Fig. \ref{fig:OneTwoFamgaugino},
\begin{eqnarray}
\Delta M_{\tilde q_{1,2},\tilde \ell_{1,2}}(\mu)\simeq \frac{g_{Y'} Y'_{\tilde q_{1,2},\tilde \ell_{1,2}}}{4\pi} M_{\tilde Z'}
\left( \ln \frac{M_{\tilde Z'}}{\mu} \right)^{3/2}.
\end{eqnarray}
This achieves our objective of raising the first two family squark and sleopton masses beyond the gaugino and the third family sfermion masses.

Since the messengers are not charged under the SM gauge group, the MSSM gaugino masses are induced only through RG running from the loops shown here (two-loop beyond the Zprimino mass), as shown in the right panel of Fig. \ref{fig:OneTwoFamgaugino},
\begin{eqnarray}
\Delta M_{\lambda_a}(\mu)\simeq \frac{g_{Y'}^2g_{a}^2 }{(16 \pi^2)^2}\, M_{\tilde Z'}
\,\left( \ln \frac{M_{\tilde Z'}}{\mu} \right).
\end{eqnarray}

The smallest is the third family members and the Higgs doublet pair. They are one more loop beyond the gaugino masses (three loop beyond the Zprimino mass)
\begin{figure}[h!]
  \begin{center}
  \begin{tabular}{c}
   \includegraphics[width=0.4\textwidth]{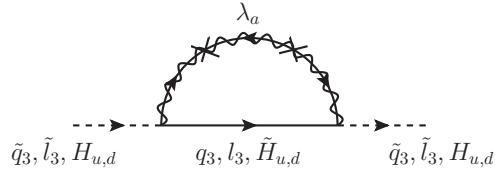}
   \end{tabular}
  \end{center}
  \caption{The mass diagrams for the third family sfermion($q_3,l_3$) and Higgs bosons. The SUSY breaking from the SM gauginos are shown as $\times$.  }
\label{fig:SMtopbottomHiggs}
\end{figure}
as shown in Fig. \ref{fig:SMtopbottomHiggs}.

The above hierarchical mass splitting is the effective SUSY scenario, through $Z'$ mediation \cite{LangZp08,JeongKimSeo}.  In the literature, it has been pointed out that if the first two family sfermions are too heavy, the stop tends to be tachyonic at the two loop level \cite{Arkstoptachyon}. At the moment, therefore, it seems that only a little hierarchy between the first two family sfermion members and those of the third family is the working one. But we speculate that some so-far unknown physics may remove this tachyonic problem as the tachyonic problem of the anomaly mediation was saved in the mirage mediation scenario.\footnote{The effective SUSY with the combined effects of gravity and $Z'$ mediations resolves this problem \cite{IeffSUSY12}.}

For an explicit calculation at the field theory level, we take $Z'=B-L$ as shown in Table \ref{table:BminusL12}.
 \begin{table}[t]
\begin{center}
\begin{tabular}{|c|cc||c|cc|}
\hline  light families &  $Y$ &$Y'$ & 3rd family and $H_{d,u}$&  $Y$ &$Y'$
\\[0.2em]
\hline &&&&&\\ [-1.1em]
$q_{1,2}$ & $\frac16$ & $\frac{1}3$ &$(t,b)$ & $\frac16$ & 0
\\[0.4em]
$u^c_{1,2}$& $\frac{-2}3$ & $\frac{-1}3$& $t^c$& $\frac{-2}3$ & 0
\\[0.4em]
$d^c_{1,2}$& $\frac{1}3$ & $\frac{-1}3$&$b^c$& $\frac{1}3$ & 0
\\[0.4em]
$l_{1,2}$& $\frac{-1}2$ & $-1$&$(\nu_\tau,\tau)$& $\frac{-1}2$ & 0
\\[0.4em]
$e^c_{1,2}$& $1$ & $1$&$\tau^c$& $1$ & 0
\\[0.4em]
$N^c_{1,2}$& $0$ & $1$& $N^c_{3}$& $0$ & 0
\\[0.4em]
& & & $H_d$& $\frac{-1}{2}$ & 0
\\[0.4em]
& & & $H_u$& $\frac{1}{2}$ & 0
\\[0.4em]
\hline
\end{tabular}
\end{center}
\caption{The $Y'$ charges of the SM fermions, Higgs doublets and heavy neutrinos.} \label{table:BminusL12}
\end{table}
\begin{figure}[t]
\begin{center}
\begin{minipage}{15cm}\hskip 3cm
  \includegraphics[width=0.5\textwidth]{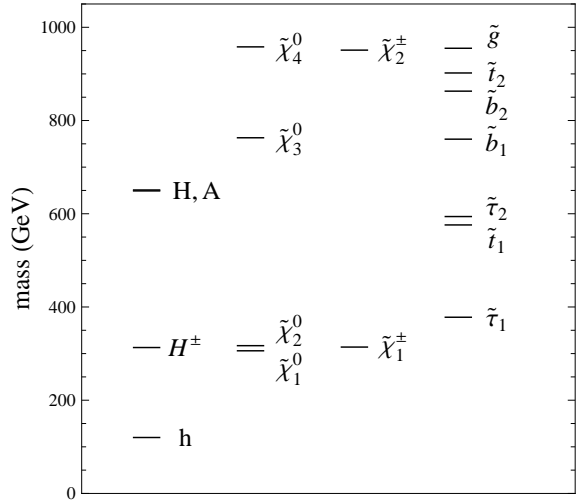}
\end{minipage}
\end{center}
\caption{The sparticle and Higgs boson mass spectra in the $Z{\,}'$ mediation.
We have taken $M_{\rm mess}=10^{14}$ GeV, $M_{Z^\prime}=10^8$ GeV and
$M_{\tilde Z^\prime}(M_{\rm mess})=1.8\times 10^6$ GeV, for which the squark masses of the first two families are above $10^6$ GeV \cite{JeongKimSeo}.}
\label{fig:SpectrumZp}
\end{figure}
The mass splitting is shown in Fig. \ref{fig:SpectrumZp}.
However, a realistic string models may give a bit different spectra from that of Fig. \ref{fig:SpectrumZp}.

In this kind of $Z'$ mediation, the LSP is most
probably the gravitino. So, CDM may be cold axions and/or axinos.
Anyway, axion seems needed also for a solution of the strong CP problem, and we proceed to comment on the axion solution of the strong CP problem.

\section{Strong CP and axions}
\vskip 0.3cm
\noindent {\bf A brief history of QCD:}

Quantum chromodynamics(QCD) started with the prediction of $\Omega^-$ \cite{Okubo62}. In terms of the current understanding, it is the completely symmetric states of three $s$ quarks and the ${\rm spin}=\frac32$ states is $s^\uparrow s^\uparrow s^\uparrow$. If $s$ quark is a fermion,  $\Omega^-$ should be completely antisymmetric under the exchange of two $s$ quarks \cite{Gursey62}. Greenberg tried a paraquark model toward this spin-statistics problem \cite{Greenberg64}. The real solution of the spin statistics problem was suggested by Han and Nambu by introducing another SU(3) degrees but with integer charged quarks \cite{HanNambu65}. Soon after the Han-Nambu paper, the fractionally charged quarks were suggested by Hori \cite{Hori66} which has been named later as QCD by Fritzsch and Gell-Mann \cite{FritGellMann}. The discovery of the asymptotic freedom crowned the QCD as the theory of strong interaction theory \cite{GrWilPol73}.

This has led to the consideration of eight gluons interacting strongly with color triplet quarks. The current quark masses were first estimated in \cite{GellMOR68} and after the advent of QCD it has been calculated again in \cite{LangPagels73}.
But, QCD with MeV order quark masses gives a light $\eta'$ \cite{Weinberg75}, which has been known as the U(1) problem. It is known that an effective interaction
of gluons $G^a_{\mu\nu}$ at the QCD $\theta$-vacuum is present,
\begin{eqnarray}
\frac{\overline{\theta}g_c^2}{64\pi^2}\epsilon^{\mu\nu\rho\sigma} G^a_{\mu\nu} G^{a}_{\rho\sigma} \label{eq:thetaterm}
\end{eqnarray}
which is a total derivative, but it cannot be neglected due to the surface contributions at Euclidian space infinity (due to instanton solutions).
Weinberg's U(1) problem is solved by this gluon anomaly term \cite{tHooft86}. This suggest that the vacuum angle term is working. Otherwise, the U(1) problem is not  solved.

This is the brief QCD history.
\vskip 0.3cm
\noindent{\bf The strong CP problem and axions:}

The anomaly term (\ref{eq:thetaterm}) breaks P and T and hence breaks the CP symmetry. It has led to the strong CP problem and three types of natural solutions \cite{KimRMP10}. Among these the axion solution which results from the spontaneously broken Peccei-Quinn(PQ) symmetry \cite{PQ77} is the beautifully realized one. The currently accepted axion solutions are the so-called {\it invisible axions} \cite{KSVZ79,DFSZ81} which live long enough to have survived until now. The axion agenda has been focussed on many different points during the last 35 years: massless up quark in 1975, the PQWW axion in 1977, the invisible axion in 1979, cosmological axions in 1983, and the axion detection experiments in 1988.

The leading contribution of the $\theta$ term is the $\eta'$ mass solving the U(1) problem. But it has led to the strong CP problem which is solved by making $\theta$ as a dynamical variable. If there is an additional almost massless dynamical field with the same anomaly term, \ie if $\theta$ is made dynamical as axion $a$, the next order contribution proportional to some light quark mass gives mass to $\theta$. This is the axion mass. Diagonalizing the $\pi^0,\eta'$ and $a$ mass, we obtain \cite{KimRMP10},
\begin{eqnarray}
m^2_{\pi^0} &\simeq& \frac{\tilde v^{\,4}}{f_{\pi^0}^2},~ m^2_{\eta'} \simeq \frac{\Lambda_{\rm inst}^4+\tilde v^{\,4}}{f_{\eta'}^2},~ m^2_{a} \simeq \frac{Z}{(1+Z)^2}\frac{f_{\pi^0}^2m_{\pi^0}^2}{F^2} (1+\Delta),\\[0.4em]
&&~{\rm with}~ \Delta=\frac{m_-}{m_+}\frac{\Lambda_{\rm inst}^4(m_+v^3+ \mu\Lambda_{\rm inst}^3)}{f_{\pi^0}^4m_{\pi^0}^4}\nonumber
\end{eqnarray}
where $m_+=m_u+m_d,\,m_-=m_d-m_u,\, Z=m_u/m_d,$ and $\tilde v^{\,4}\equiv m_+v^3+ 2\mu\Lambda_{\rm inst}^3$.

The dark matter particle in the universe is the most looked-for one(s) in cosmology and at the LHC, and also at low temperature labs: at the mass scales of 100 GeV (for WIMP) and  10--1,000 $\mu\,$eV (for axion). The axion potential is almost flat for a large axion decay constant $F$. In this axion potential, the minimum point $\theta=0$ is the CP conserving point. In the evolving universe, at some temperature, say $T_1$, $a$ starts to roll down the hill to the CP conserving point $\theta=0$. This analysis depends on the axion decay constant and the initial VEV (the so-called misalignment angle) of $a$ at $T_1$. A recent analysis of these effects has been given in \cite{BaeHuh08}.

Fortunately, the invisible axion is at the level of possible detection through $\gamma-\gamma- a$ and/or $\bar e-e-a$ interaction. By discovering the axion, QCD is completed satisfactorily. If the axion is ruled out at the axion window, either its mass is too small to be detected, or the fine-tuning problem of $\theta$ remains unsolved theoretically.

\vskip 0.3cm
\noindent {\bf A new form of the CKM matrix:}

\begin{figure}[!t]
  \begin{center}
  \begin{tabular}{c}
   \includegraphics[width=0.45\textwidth]{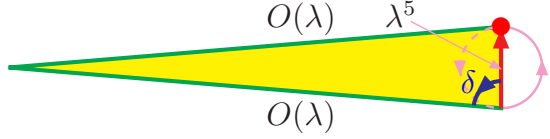}
   \end{tabular}
  \end{center}
 \caption{The Jarlskog triangle. This triangle is for two long sides of $O(\lambda)$. Rotating the $O(\lambda^5)$ side (the red arrow), the area becomes maximum as the CP phase $\delta$ becomes approximately $\frac{\pi}2$ \cite{KimSeoCKM}.
  }
\label{fig:JTriangle}
\end{figure}

I have discussed some issues related to SUSY and axion. Related to axion, I discussed
a brief history of QCD, the need for a solution of the strong CP problem, and the
invisible axion. Since  the discussions here are related to CP, a good weak CP violation parametrization may be useful. One recent CKM parametrization allows one to look at the Jarlskog triangle from the CKM matrix itself \cite{KimSeoCKM}. Studying in this new parametrization, the maximum CP violation in the quark sector is apparent as shown in Fig. \ref{fig:JTriangle}.
In the recent study \cite{KimSeoCKM}, we showed that by making the whole determinant of the CKM matrix $V$ real, the imaginary part of any one term of the determinant of $V$ (e. g. $|{\rm Im}\,V_{31}V_{22}V_{13}|$) is the Jarlskog determinant $J$. This method of calculating $J$ is much easier than calculating \cite{Jarlskog85}
$$
J = \frac{-{\rm Det.}\, C}{2F(m_{t,c,u})F(m_{b,s,d})}
$$
where $F(m_{t,c,u})=(m_t-m_c)(m_t-m_u)(m_c-m_u)$ and $F(m_{b, s,d})=(m_b-m_s)(m_b-m_d)(m_s-m_d)$, and $C$ is
$iC=[M_u, M_d]$,
where $M_{u,d}=L^{(u,d) \dagger}M^{0(u,d)} L^{(u,d)}$ with the diagonal $M^{0(u,d)}$. It can be also written as $iC=[M_u M_u^\dagger, M_d M_d^\dagger].$
With this method, it is easy to convince oneself that the weak CP violation data shows the maximality as shown in Fig. \ref{fig:JTriangle}.

\section{Higgs boson at 125 GeV ?}

It seems that some new particle is present around 125 GeV. Most probably, it looks like a Higgs boson. Is it the SM Higgs boson or the lightest or the next lightest scalar Higgs boson of the NMSSM, or something else?

In SUSY models, the probability for it to be the MSSM Higgs boson is low \cite{Earler12}, and a singlet $X$ is introduced to fit the LHC preliminary data. These SUSY models with one more singlet is next MSSM(NMSSM). Supergravity models with one more singlet $X$ is quite plausible, in particular in string compactifications toward SUSY standard models \cite{SSMkimjh07}. Here, we point out that the pseudoscalar in $X$ can be at the electroweak scale (even if it is not the 125 GeV particle) and its two photon decay mode is discussed here.

We require that the strong CP problem is solved in this NMSSM. So, the PQ symmetry in the NMSSM is needed, which we call N$_{\rm PQ}$MSSM. The MSSM cannot achieve this because $\mu H_uH_d$ in the MSSM breaks the PQ symmetry. Then, the question is, ``What is the spectrum of the N$_{\rm PQ}$MSSM?" In Refs. \cite{KNS12} and \cite{Yamaguchi11},  N$_{\rm PQ}$MSSM has been discussed. The axino mass in the N$_{\rm PQ}$MSSM has been discussed in \cite{KimSeo12} where the heavy axino is most likely in contrast to the light axino study of \cite{AxinoGrav94}. The light pseudoscalar in the  N$_{\rm PQ}$MSSM has been emphasized in \cite{KNS12}.

\begin{eqnarray}
W=-H_uH_dX +mX\overline{X} -\eta \overline{X}\,S_1^2-\xi H_uH_dX' +m' X'\, \overline{X} +Z_1(S_1S_2-F_1^2)+Z_2(S_1S_2-F_2^2) \label{eq:Whigh}
\end{eqnarray}
where the $Z_1$ and $Z_2$ terms are introduced to parametrize SUSY breaking and the PQ symmetry breaking \cite{Kim84}. $Z_1$ and $Z_2$ of (\ref{eq:Whigh}) is of order TeV but we assume them sequestered from the other light fields since they are assumed to mimick the DSB scheme. Beyond $Z_1$ and $Z_2$, there can be electroweak scale singlets. If we introduce just one, say $X_{\rm ew}$, the electroweak scale superpotential from the  N$_{\rm PQ}$MSSM  is
\begin{eqnarray}
W_{\rm ew}=-\mu H_uH_d  -f_h H_uH_d  X_{\rm ew}+\cdots \label{eq:Wlow}
\end{eqnarray}
For example, $X_{\rm ew},\, X_{\rm ew}^2$ and $X_{\rm ew}^3 $ terms cannot be present \cite{KNS12,KimSeo12}. How $X_{\rm ew}$ survives down to the electroweak scale depends on the details of the scheme. In fact, finding a reasonable scheme is pain in the neck. At the electroweak scale, we use the $\mu$ term  generated at the high energy scale \cite{KimNilles}, and also the $H_uH_d X_{\rm ew}$ renormalizable term.  The important thing is that $X_{\rm ew}$ survives down to the TeV scale, which we do not question here. Treating $\mu$ as a perturbation, there exists a symmetry:
\begin{eqnarray}
X_{\rm ew}\to e^{-2i\delta}X_{\rm ew}, ~ H_u\to e^{+i\delta}H_u, ~ H_d\to e^{+i\delta}H_d.\label{eq:KNSsymm}
\end{eqnarray}
Then, the chargino (more generally the Higgsino pair) has the anomaly current, and there results the pseudoscalar $a_{_X{_{\rm ew}}}$ coupling to the anomaly term of the electromagnetic field, $a_{_X{_{\rm ew}}}F_{\mu\nu}^{\rm em}\, \tilde F_{\rm em}^{\mu\nu}$. The symmetry (\ref{eq:KNSsymm}) originating from the PQ symmetry gives
a light pseudoscalar if $\mu\simeq 0$. The photon (or weak gauge boson) fusion gives the two photon mode which can be detected in the future.

\section{Conclusion}
In view of the recent discovery of the Higgs boson, we paid attention to some effective supersymmetric models, \ie the light Higgs boson around 125\,GeV is attempted to be accommodated. A note on an extra $Z'$ from GUTs, and the scalar and pseudoscalar Higgs bosons in supersymmetric extension of the standard model are also discussed. We also discussed a few topics related to QCD such as the strong CP problem and the recent parametrization of the weak CP violation where the physical quantity related to the Jarlskog triangle is easy to identify.

\begin{figure}[!t]
  \begin{center}
  \begin{tabular}{l}
  {\includegraphics[width=0.9\textwidth]{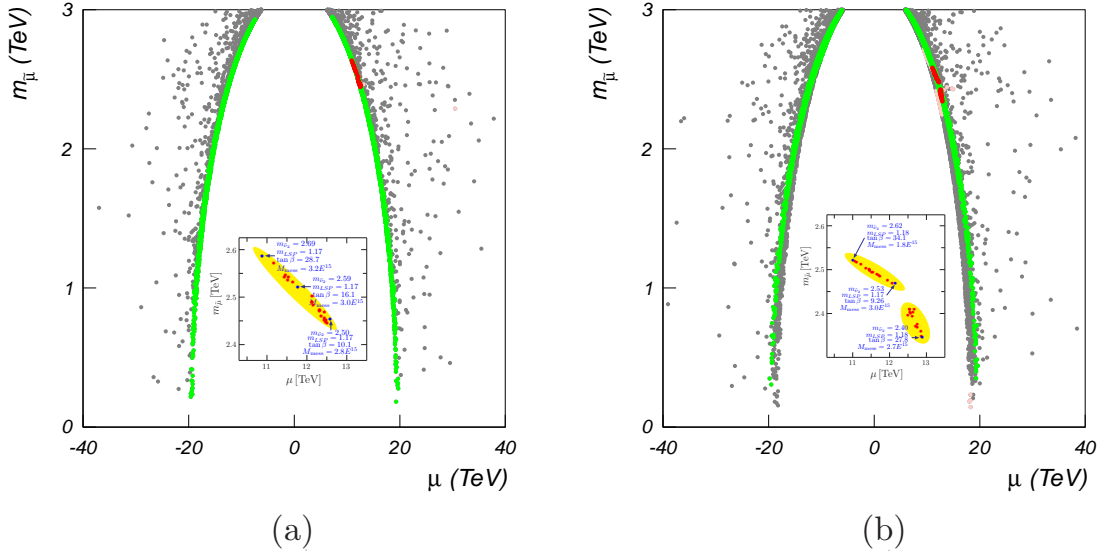}}
       \end{tabular}
\caption{
The scatter plot in the $m_{\tilde\mu}-\mu$ space for $\Lambda_h=3\times 10^{13}\gev$ out of $10^5$ trial points: (a) for  $\lambda_f=\frac16,\,\lambda_h=0$, and (b) for  $\lambda_f=-1,\,\lambda_h=0$ with the notation of Ref. \cite{IeffSUSY12}. The scanned parameters are $\mu$ and $M_{\rm mess}=(0.1\sim 10)\times 10^{15}\,\gev$. The top quark mass corresponds to $m_t=(173.5\pm 1.4)\,\gev$.  The gray dots are the trial points.  The green dots are those satisfying the LHC constraints ($m_{\tilde q_{1,2}}>1.5\,\tev$ and the LHC gluino mass bound) and $m_h=(125\sim 127)\,\gev$. The pink dots are filtered by $(g-2)_\mu$. The red dots are those satisfying all the constraints including  $(g-2)_\mu$. In the enlarged insets, some selected red points are shown again with more information in blue dots .
}\label{fig:ConstraintsSmu}
  \end{center}
\end{figure}

\vskip 0.3cm
[Note added after the talk: In this talk I presented my recent works related to the effSUSY. But after the Higgs boson discovery, it became urgent to interpret the 3.6 $\sigma$ data on $(g-2)_\mu$ of the Brookhaven National Laboratory(BNL) \cite{gmtwoBNL}. The 125 Higgs boson with the BNL $(g-2)_\mu$ is attempted to be explained in the MSSM framework with the inverted effSUSY(IeffSUSY) \cite{IeffSUSY12}. The related parameters in the IeffSUSY are shown in Fig. \ref{fig:ConstraintsSmu}. This figure is based on the $Z'$ and gravity mediations with the third family sfermions are the heaviest, which is possible with the $Z'$ quantum numbers of Table \ref{table:BmL}.]

\begin{table}[!h]
\begin{center}
\begin{tabular}{|ccc|ccc|ccc|}
\hline  1$^{\rm st}$ f.  & $Y$ &$Z'$ & 2$^{\rm nd}$ f. &  $Y$ &$Z'$ &  3$^{\rm rd}$ f.  &  $Y$ &$Z'$
\\[0.2em]
\hline &&&&&&&& \\ [-1.1em]
 $(u,d)$ & $\frac16$ & $\frac{1}3\lambda_f$ & $(c,s)$ & $\frac16$ &   $\frac13 \lambda_f$ & $(t,b)$ & $\frac16$ & $\frac13 $
\\[0.4em]
$u^c $& $\frac{-2}3$ & $\frac{-1}3\lambda_f$ & $c^c$& $\frac{-2}3$ &  $\frac{-1}3 \lambda_f$ & $t^c$& $\frac{-2}3 $ &$\frac{-1}3$
\\[0.4em]
$d^c $& $\frac{1}3$ & $\frac{-1}3\lambda_f$ &$s^c$ & $\frac{1}3$ & $\frac{-1}3 \lambda_f$ &$b^c$ & $\frac{1}3$  &  $\frac{-1}3$
\\[0.4em]
 $(\nu_e,e)$& $\frac{-1}2$ & $-\lambda_f$ &$(\nu_\mu,\mu)$ & $\frac{-1}2$ & $0$&$(\nu_\tau,\tau)$& $\frac{-1}2$ & $-(1+\lambda_f)$
\\[0.4em]
$e^c$ & $1$ & $\lambda_f$ & $\mu^c$ & $1$ & $0$ & $\tau^c$& $1$ & $1+\lambda_f$
\\[0.4em]
$N^c_{1}$& $0$ & $\lambda_f$ & $N^c_{2}$ & $0$ & $0$& $N^c_{3}$& $0$ & $1+\lambda_f$
\\[0.4em]
\hline
\end{tabular}\vskip 0.3cm
\begin{tabular}{|ccc|}
\hline  $H_{d,u}$& $Y$ &$Z'$
\\[0.2em]
\hline  &&\\ [-1.1em]
 $H_d$ & $\frac{-1}{2}$ & $\frac{-1}{2}\lambda_h$
\\[0.4em]
 $H_u$ & $\frac{1}{2}$ & $\frac{1}{2}\lambda_h$
\\ [0.2em]
\hline
\end{tabular}
\end{center}
\caption{The $Z'$ charges of the SM fermions, Higgs doublets and heavy neutrinos. The choice $\lambda_f=\lambda_h=0$ is the simplest case. For the BNL $(g-2)_\mu$, the muonic leptons carry vanishing $Z'$ charges while the others carry nonzero $Z'$ charges with nonzero $\lambda_f$.} \label{table:BmL}
\end{table}

\vskip 0.5cm
  This work is supported in part by the National Research Foundation (NRF) grant funded by the Korean Government (MEST) (No. 2005-0093841).

\vskip 0.5cm

\end{document}